# " Music-ripping " : des pratiques qui provoquent la musicologie

**Francis Rousseaux**
francis.rousseaux@ircam.fr
Equipe IST CUIDADO
IRCAM-CNRS
1, place Igor-Stravinsky
75004 Paris
France
Tél. : 00 33-1-44-78-48-39

**Alain Bonardi**
alain.bonardi@wanadoo.fr
Maison des Sciences de l'Homme Paris Nord
4, rue de la Croix Faron
93210 Saint-Denis La Plaine
France
Tél. : 00 33-1-55-93-93-00

**Résumé**

Hors des placements habituels de l'informatique dans le domaine de la musique et de la musicologie, on constate depuis peu l'apparition de systèmes hommes-machines qui manifestent leur existence en rupture avec une tradition certes encore jeune, mais déjà opérante.

Car si ces systèmes singuliers opèrent incontestablement dans les champs d'expansion usuels de la musique, ils ne font aucune référence systématique aux catégories musicologiques connues. Au contraire, les expérimentations qu'ils rendent possibles inaugurent des usages où l'écoute, la composition et la transmission musicales se confondent dans un geste parfois qualifié de " music-ripping ".

Nous montrerons en quoi les pratiques de " music-ripping " provoquent la musicologie traditionnelle, dont les catégories canoniques s'avèrent ici impuissantes au compte-rendu.

Pour ce faire, il nous faudra :
– expliciter un jeu de catégories minimal qui suffise à sous-tendre les modèles usuels de la musique assistée par ordinateur;
– faire de même pour les systèmes hommes-machines (anti-musicologiques ?) dont l'existence nous trouble;
– examiner les conditions de possibilité de réduction du second ensemble catégorial au premier;
– conclure sur la nature du " music-ripping ".

**Abstract**

Out of the scope of the usual positions of computing in the field of music and musicology, one notices the emergence of human-computer systems that do exist by breaking off.

Though these singular systems take effect in the usual fields of expansion of music, they do not make any systematic reference to known musicological categories. On the contrary, they make possible experiments that open uses where listening, composition and musical transmission get merged in a gesture sometimes named as " music-ripping ".

We will show in which way the music-ripping practices provoke traditional musicology, whose canonical categories happen to be ineffectual to explain here.

To achieve that purpose, we shall need:
– to make explicit a minimal set of categories that is sufficient to underlie the usual models of computer assisted music;
– to do the same for human-computer systems (anti-musicological?) that disturb us;
– to examine the possibility conditions of reduction of the second set to the first;
– to conclude on the nature of music-ripping.

# 1. Table d'orientation des positionnements de l'informatique dans les pratiques musicales et musicologiques

Comment l'informatique se positionne-t-elle dans les pratiques musicales et musicologiques ? Partant de nos expériences et observations depuis plusieurs années, nous proposons ici une table d'orientation. Elle dépasse le cadre de l'informatique musicale, essentiellement considérée dans sa destination aux compositeurs et créateurs depuis ses débuts il y a une quarantaine d'années.

Nous observons trois positionnements typiques qui ne recouvrent aucunement la " partition " fonctionnelle des pratiques musicales et musicologiques :
- l'outillage informatique des pratiques traditionnelles;
- le placement de l'informatique comme substrat universel de la musique;
- la création de dispositifs informatisés pour interroger les pratiques habituelles.

Face à ce dispositif vient en rupture :
- la réalisation de systèmes hommes-machines singuliers qui n'invoquent pas la musicologie.

Bien entendu, ces quatre positionnements stylisent et discrétisent un continuum, une réalité qui se glisse souvent entre leurs marques.

## 1.1. Quand l'ordinateur outille les pratiques traditionnelles

Ce positionnement de l'informatique est le plus répandu : telle ou telle activité auparavant exercée par des intervenants humains est " réorganisée " en y associant des ordinateurs, dont on exploite habilement les capacités de traitement et de stockage. Efficace et productive, cette approche permet de faire partager des éléments en rapport avec la musique (connaissances historiques, théoriques ou documents d'archives) sous la forme de fichiers textuels, musicaux, graphiques ou vidéo et de traitements centrés sur l'extraction de données, avec plus ou moins de raffinement.

*Exemples :*
- les fiches sur les œuvres musicales dans les centres de ressources deviennent des bases de données, qui facilitent le travail d'investigation du musicologue en permettant des recoupements d'informations. Ainsi, la base de données de la Médiathèque de l'Ircam autorise la recherche des références par effectif instrumental ou par date de composition ou de création [Fingerhut 1999];
- L'ordinateur sert de professeur d'écriture musicale à distance par Internet, capable de proposer et corriger des exercices d'harmonie tonale [Jouvelot 1998].

## 1.2. Quand la musique donne à entendre l'informatique

Renversant le point de vue précédent, plusieurs approches ont utilisé les fortes capacités combinatoires symboliques de l'ordinateur pour produire de la musique, en proposant des algorithmes de génération. L'intérêt de ces approches vient de leur capacité à donner à entendre des structures qui ne sont pas nativement musicales.

*Exemples :*
- Les grammaires musicales génératives de Lerdahl et Jackendorff [Lerdahl & Jackendorff 1985] proposent une approche directement tirée des travaux de Chomsky en linguistique;
- Dans de nombreuses compositions, Iannis Xenakis [Xenakis 1981] met en œuvre des techniques probabilistes variées (stochastiques, markoviennes) que les musiciens doivent rendre, faisant souvent face à un excès du modèle par rapport à ce qui peut être joué, et donc à une nécessité de filtrer musicalement cette matière première, comme en témoigne le pianiste Claude Helffer [Albèra 1995].

### 1.3. Quand des dispositifs informatisés interrogent les pratiques

A côté des deux positionnements précédents en tête-bêche, des concepteurs de cédéroms créatifs ont proposé au cours des années 1990 des dispositifs originaux qui interrogent la tradition-même de l'interprétation, de l'analyse et de la composition.
*Exemples :*
- Le cédérom *Prisma* consacré à la compositrice Kaija Saariaho[1], envisage la composition comme un jeu interactif. La pièce *Mirrors* pour flûte et violoncelle a été découpée en courts fragments que l'utilisateur peut recombiner librement ou en suivant des règles compositionnelles;
- Le cédérom *Les musicographies* conçu par Dominique Besson[2] et réalisé par Olivier Koechlin propose de faire " sonner " des représentations graphiques de sons, qui questionnent aussi les modalités littéraires de l'analyse musicale.

### 1.4. Quand naissent de singuliers systèmes hommes-machines " a–musicologiques "

Ces dernières années ont vu la réalisation de systèmes musicaux qui n'invoquent pas systématiquement la musicologie. Les catégories musicologiques en sont absentes en tant que telles, seules quelques notions sont empruntées à d'éphémères occasions. C'est pourquoi nous les qualifions provisoirement " d'a-musicologiques ", car ils ne convoquent pas les moyens habituels de parler de la musique.
Exemples :
- L'opéra interactif *Virtualis* : dans ce projet d'opéra numérique sur support cédérom [Bonardi & Rousseaux 2002], l'utilisateur est amené à naviguer dans un espace tri-dimensionnel et à transformer des contenus sonores spatialisés en manipulant leurs représentations graphiques (polygones, surface, etc); à aucun moment le visiteur n'est confronté aux catégories musicologiques : le calcul des variations de fragments se fait par simple mouvement de souris dans l'espace 3D;

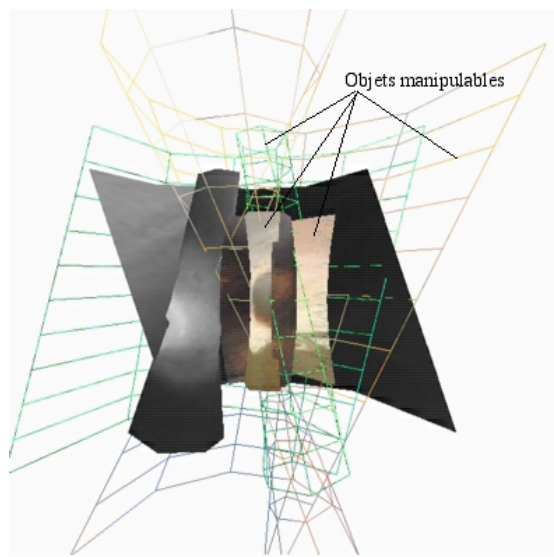

**Figure 1.** *Exemple d'objets 3D associés à des contenus musicaux dans Virtualis (source : Alain Bonardi)*

---

[1] Cédérom produit en 1999 par le Finnish Music Information Centre.
[2] Cédérom édité par l'INA-GRM en 1995.

- Les systèmes de partage de fichiers musicaux au format MP3 fondés sur l'architecture en réseau " peer-to-peer " comme le célèbre site Napster : les utilisateurs de ces systèmes téléchargent des morceaux qu'ils ne cessent d'accumuler et de réorganiser sur leur propre disque dur, en les séquençant, en les mixant, en construisant des compilations. Souvent ces fichiers sont à leur tour copiés et recompilés vers des baladeurs à disque dur pour une écoute dite " nomade ". Comme le remarque François Pachet[3], " une bonne partie de mon rapport à la musique numérique consiste à contempler les noms de fichiers, à projeter des idées de noms sur eux, à en inférer des informations. "

La conception de ces systèmes les exclut des positionnements précédents, car ils posent très clairement qu'entendre et manipuler participe de la même chose. De ce fait, si phénoménologiquement, ils se donnent comme " a-musicologiques ", rien n'empêche de les considérer comme musicologiques dans le sens très vaste où ils opèrent dans la transmission de la musique.

Mais pour l'instant nous ne disposons que d'un ensemble d'exemples regroupés sous la bannière " a-musicologique ", figurant comme reliquat de ce que les trois placements typiques n'ont pu absorber. Nous devons pourtant mettre en relation cet ensemble *a priori* disparate avec les catégories musicologiques traditionnelles, pour préciser son rapport problématique à la musicologie.

## 2. Etude des systèmes hommes-machines " a–musicologiques "

Pour établir la commensurabilité de cet ensemble de systèmes hommes-machines " a-musicologiques " avec les catégories de la musicologie, nous devons modéliser les deux versants à comparer.
Nous proposons pour chaque versant un modèle à la fois rustique et perfectible :
- un modèle de la musicologie au sens habituel;
- un modèle des systèmes hommes-machines " a-musicologiques ", qui nous permettra le cas échéant de catégoriser notre ensemble d'exemples.

Puis nous tenterons d'établir un morphisme entre les deux modèles. Trois cas sont alors possibles :
- soit un tel morphisme est exhibé entre les deux modèles, modèles qu'il conviendrait alors d'affiner pour préciser le morphisme;
- soit on démontre qu'il n'existe pas de morphisme entre les deux modèles et dans ce cas le second modèle, irréductible, deviendrait " anti-musicologique ", en ce sens ad hoc;
- soit on ne parvient à aucun des deux résultats précédents; dans ce cas, le second modèle conserverait par défaut sa réputation " a-musicologique " ainsi que nous l'avions qualifié au départ.

### 2.1. Modèle robuste de la musicologie

Nous l'organisons en quatre pôles-attracteurs. Les trois premiers sont familiers aux musicologues, le quatrième mérite d'être précisé et illustré.
- **l'ensemble des corpus historiques et sociologiques** proposant une compréhension globale du phénomène musical, de l'entour multiple des créateurs, des interprètes, des spectateurs et des œuvres;

---

[3] Extrait d'un texte à paraître intitulé " Nom de fichiers : le nom – Tentative phénoménologique de description de mon activité d'écoute musicale à partir de supports numériques – Ecouter de la musique/regarder des fichiers, communiqué personnellement par l'auteur.

- **l'ensemble des modèles de l'individualité de l'artiste " inspiré "** en tant que compositeur ou interprète : psychologique, psychanalytique, hagiographique, sans oublier l'étude " chorégraphique " du geste musical;
- **l'organologie** et plus largement les phénomènes acoustiques, c'est-à-dire l'étude de tout ce que la musique doit à la lutherie et aux conditions et lieux d'écoute et de création;
- **la raison graphique**, c'est-à-dire tout ce qui ne se comprend que dans l'inscription bidimensionnelle de la musique sur support. Ainsi, les théories musicales allant de la constitution des échelles jusqu'à l'orchestration en passant par l'harmonie et le contrepoint, et se donnant comme " jurisprudence " les grandes partitions-modèles du passé, sont-elles fondées sur l'inscription orthogonale des hauteurs (notation proportionnelle : plus la note est aiguë, plus il faudra de lignes supplémentaires) et des durées (notation algébrique, dans laquelle des entités abstraites ont des rapports figés entre elles : ainsi dit-on qu'une noire vaut deux croches). Il y a donc là une raison graphique, qui s'impose comme fondement de toute théorie musicale, au sens de Goody [Goody 1987]. Ainsi Hugues Dufourt [Dufourt 2002] soutient-il que la musique grecque antique, à l'image de sa pensée philosophique, cherche " l'un dans le multiple, l'immuable dans le changement ", alors que la polyphonie occidentale et la mentalité moderne " demande[nt] le déploiement du temps, puis de l'espace ". Cette différence se saisit parfaitement à la lumière des écritures musicales : l'écriture grecque antique, uniquement un alphabet de hauteurs, nie le temps, et ne fait qu'occuper son support, alors que l'écriture occidentale prend la pleine mesure de son support-plan et de structures de pensées bi-dimensionnelles ouvrant à la polyphonie.

Peut-on réduire la musicologie à ces quatre pôles-attracteurs ?
Si nous nous restreignons au champ de l'analyse musicale, il est assez aisé de rendre compte des différentes méthodes analytiques en combinant ces quatre pôles. Ainsi, les fondements d'un plan tonal sont à chercher du côté de l'acoustique et du côté de la raison graphique. Une méthode comme l'analyse narrative [Grabocz 1999] est plus difficile à situer car faisant appel aux quatre pôles à travers la notion de style ou topique. Mais il faut reconnaître que le passage par le corps, comme exigence d'une compréhension de l'interprétation qui n'exclut pas la corporéité [Rink 2002], s'intègre mal au modèle robuste.

Reste que ce modèle permet une première appréhension assez efficace de la musicologie.

### 2.2. Modèle robuste de l'ensemble des systèmes hommes-machines " a–musicologiques "

Nous posons quatre pôles-attracteurs, déduits directement de nos expériences :

- **le calcul algorithmique** : l'approche par le calcul est fondée sur l'évaluation de l'état courant de la science et des techniques de calcul (algorithmique), ainsi que sur des intuitions de débouchés industriels. S'agissant du son et de la musique, la question soulevée est celle-ci : peut-on trouver des algorithmes de calcul qui s'appliqueraient au signal numérique pour en extraire des valeurs de ce qu'on pourrait alors appeler des attributs descripteurs de ce signal, qu'il s'agirait alors de baptiser de façon à se souvenir du rôle heuristique central qu'ils peuvent jouer pour représenter le signal pour des applications particulières ? Un descripteur est donné par un nom de baptême, un algorithme d'extraction, et une liste d'applications potentielles susceptibles d'en valoriser l'usage sur des marchés. Cette approche est typique des pratiques du consortium " Motion Picture Expert Group " (MPEG).

*Exemple :*
La " durée effective d'un signal sonore " constitue un descripteur calculable à partir d'un enregistrement : il correspond à l'évaluation de la durée pendant laquelle le signal est significatif au plan perceptif. Cette durée est calculée sur la base d'un seuil d'enveloppe d'énergie obtenu à partir d'études psycho-acoustiques. Elle permet par exemple de discriminer un son percussif d'un son entretenu. Cet attribut sera probablement intégré par le consortium de normalisation MPEG7.

- **les réseaux de connaissances** sont le résultat de phases d'acquisition de connaissances dans un domaine donné, conduisant à la spécification d'un vocabulaire le représentant, qui est encodé sur ordinateur principalement sous forme de classes, de structures et de fonctions. La contextualisation, qui est l'une des difficultés principales dans la modélisation de la pensée, est résolue par le traitement logique de ces unités de sens réifiées. La mise en œuvre de réseaux de connaissances donne à croire que les objets de pensée sont disponibles sur étagère et ouvre la possibilité de faire sens à partir de la confrontation artificielle de telles structures.

    *Exemples :*
    Un réseau de connaissances dans le domaine des objets sonores est la description des sons sur le mode de l'écoute réduite, proposée par Pierre Schaeffer dans son *Traité des objets musicaux* [Schaeffer 1966] et reprise par Michel Chion [Chion 1983]. Dans le domaine de l'informatique musicale, nous pouvons citer le système MusES [Pachet 1994] qui contient une représentation par objets des concepts de base de la musique tonale et de leurs propriétés principales (notes, pitch-classes, gammes, intervalles, accords, mélodies, etc.).

- **le rapport charnel à la perception et à la compréhension réconciliées** : dans un contexte musical, face à un ordinateur, l'utilisateur agit tout en téléchargeant des sons : en reprenant le slogan publicitaire d'une célèbre marque d'ordinateur (" rip music ") on peut parler de " music-ripping " pratiqué par des " music-rippers " qui attrapent et éventrent[4] des contenus musicaux. C'est une esthétique de la retouche permanente, plus proche en cela de la poterie où il faut continûment affiner l'objet que de la sculpture travaillée par coups et éclats.

    *Remarque :*
    Notre éventreur de musique ne s'en tient pas à la capture et à la déformation. Il n'a de cesse de transmettre ses fichiers à d'autres " music-rippers " par email, par téléchargement, par l'organisation de sessions sur Internet, en encore par des réseaux " peer-to-peer ". Ce que transmet le " music-ripper " est une inscription de son écoute continûment façonnée; son rôle devient comparable à celui des transcripteurs dans les siècles passés, qui inscrivaient une écoute d'une symphonie ou d'un opéra dans une partition de piano [Szendy 2001].

- **la raison computationnelle** : les ordinateurs comme outils de calcul ont la double capacité de supporter des traitements temps réel et de simuler la représentation de connaissances utiles au raisonnement et à l'organisation. Mais cette double aptitude ne s'exerce pleinement que par leur faculté de répétition. Elle les rend certes capables de reproduire fidèlement des résultats calculés, mais surtout participe de la simulation comportementale, en soumettant des processus exécutés à l'investissement d'esprit des utilisateurs humains. L'ordinateur répète, et cela provoque l'utilisateur, qui souvent sollicite cette répétition pour tenter de définir un modèle, c'est-à-dire établir par les figures du discernement et de la différence ce qui constitue le Même. Il forme ainsi des représentations bien vite ressaisies à leur tour par l'interprétation.

---

[4] Le slogan " rip music " désigne explicitement cette activité d'appropriation/ré-appropriation musicale. Dans la connaissance commune de la langue anglaise, le verbe " to rip " a été popularisé par la personnalité de " Jack The Ripper " autrement dit Jack l'Eventreur (qui connut d'ailleurs une certaine destinée musicale, puisqu'il apparaît à la fin de *Lulu* de Berg).

*Exemple :*
Ecouter, c'est vouloir écouter encore : on désire un prolongement de l'expérience. Mais aussi paradoxalement vouloir écouter encore autre chose : on désire un nouvel objet d'expérience pour que l'expérience persiste. Cette écoute idéale se décline en la construction d'une séquence musicale s'appuyant sur une affinité élective toujours critique. Grâce aux enregistrements musicaux, mobilisables sans délai par le biais de systèmes d'accès et de dispositifs de restitution, écouter signifie composer une séquence. Les fonctions d'accès direct aux plages musicales permettent une répétition de l'écoute, pour différencier et regrouper des morceaux choisis, par exemple sous forme d'une " playlist " (ou séquence favorite) de fichiers musicaux.

Le tableau ci-après établit une synthèse des éléments relatifs à chacun des quatre pôles que nous venons d'exposer. Nous y avons de plus inclus une autre catégorie importante, que nous désignons sous le nom " d'ordres pertinents ". Il s'agit non seulement de comprendre les pôles en tant qu'ensembles, mais aussi de saisir ce qui permet à l'intérieur de chacun d'eux d'ordonner les éléments. Ainsi, les descripteurs du calcul algorithmique peuvent être ordonnés de deux manières : par les relations de dépendance propres au calcul (la valeur d'un descripteur n'est donnée qu'après évaluation d'un autre), ou selon la rentabilité économique envisagée.

### *2.3. Etude de la commensurabilité entre les deux modèles*

Nous avons exhibé deux modèles à quatre pôles-attracteurs :

- d'une part, l'ensemble des corpus historiques et sociologiques, l'ensemble des modèles de l'individualité de l'artiste " inspiré ", l'organologie, et la raison graphique;
- d'autre part, le calcul algorithmique, les réseaux de connaissances, le rapport charnel à la perception et la compréhension, ainsi que la raison computationnelle.

Comment les comparer et chercher des morphismes les reliant ? Cette question constitue en soi un véritable programme de recherche, qui dépasse le cadre de cet article.

| Approche | Calcul | Réseaux de connaissances | Rapport charnel à la perception et à la compréhension | Raison computationnelle |
| --- | --- | --- | --- | --- |
| Signe | Excès de **l'inscription** | Médiation de **la classification** | Fusion entre **écoute** et **manipulation** | Investissement d'esprit des processus de **répétition** machinique |
| Modalité | Calcul sur le signal numérique | Représentation des connaissances | Pratique corporelle d'une écoute/création et de sa transmission | Simulation comportementale |
| Ordres pertinents | Deux ordres sur les descripteurs : <br> - un ordre de dépendance du calcul <br> - un ordre de rentabilité économique | Acquérir les connaissances et les ordonner sont les deux faces d'une même activité de modélisation, qui a bien du mal à choisir entre des ordres théoriques (l'abstraction ou la généralisation) et des ordres pratiques (utilité pour un usager), d'où des problèmes d'acquisition et de maintenance des ontologies | Deux ordres pour classer les systèmes : <br> - potentiel de transformation offert par les contenus et les logiciels " players " <br> - degré d'interoperabilité des supports technologiques | Deux ordres pour classer les situations : <br> - la confiance de l'utilisateur <br> - la productivité des systèmes homme-machine |
| Lieux de développement | Le consortium de normalisation MPEG | Les communautés universitaires se réclamant de " l'acquisition des connaissances " | Les réseaux de partage/piratage de fichiers musicaux | Les laboratoires de conception des baladeurs musicaux numériques à disque dur |
| Exemple musical | La " durée effective " d'un signal sonore est un descripteur MPEG en voie de normalisation | Dans son *Traité des objets musicaux*, Pierre Schaeffer a proposé une ontologie pour la description des sons sur le mode de l'écoute réduite (au sens de la phénoménologie husserlienne) | Le " music-ripping " | Les logiciels de confection de " playlists " |

Une première direction de recherche serait l'étude de relations entre certains pôles, et plus particulièrement entre le pôle " raison graphique " dans le premier modèle et le pôle " raison computationnelle " dans le second, à la lumière des écrits de Rastier [Rastier 1994] et Bachimont. Pour amorcer la réflexion, disons que nous avons remarqué à quel point la raison computationnelle dépend de la capacité des ordinateurs à la répétition, permettant le discernement et la différenciation. En revanche, la raison graphique ne s'appuie pas sur la répétition de l'expérience d'investissement d'esprit, elle propose d'emblée des catégories différenciées. Ainsi, pour prendre un exemple musical, le traitement d'un son dans un logiciel d'informatique musicale repose sur l'écoute répétée de fragments-clés (le plus souvent des transitions) alors qu'un arrangement musical d'une composition en partition-papier passe toujours par l'identification de structures directement interprétables sur la partition.

Une deuxième direction de recherche serait la confrontation des deux modèles à l'aune d'une catégorie tierce. Nous pensons ici à une catégorie ayant *a priori* peu de rapports avec notre question, celle de la *représentation,* par exemple telle qu'elle fut abordée lors des dernières séances du Séminaire InterArts de Paris, coordonnée par Danièle Pistone : si l'on peut opposer la représentation comme jeu d'artiste et d'exégète dans le spectacle vivant traditionnel à la représentation comme condition de possibilité de l'interaction homme-machine dans l'art numérique, il est peut-être possible de différencier nos deux modèles sur des bases analogues, notamment en remarquant que le second impose ses formalismes de manipulation, alors que le premier tente de représenter des situations musicologiques qui s'imposent sans prescrire ces codes d'action vis-à-vis des machines.

**Conclusion**

Nous avons présenté les trois positionnements classiques de l'informatique par rapport aux pratiques musicales et musicologiques : outillage des pratiques traditionnelles par l'ordinateur, détournement de structures générées non nativement musicales, création de dispositifs interrogeant les pratiques habituelles. Ils invoquent tous systématiquement les catégories de la musicologie en tant que telles.
Ce n'est pas le cas des systèmes hommes-machines singuliers, bien qu'opérant dans le champ de la musique, qui font leur apparition depuis quelques années. Ecoute et manipulation y sont confondues dans un geste parfois qualifié de " music-ripping ".

Nous avons proposé deux modèles comprenant chacun quatre pôles-attracteurs, 1° des catégories de la musicologie et 2° de l'ensemble de systèmes hommes-machines singuliers. Demeure la difficile question du caractère " a-musicologique " de ces systèmes : la comparaison fine des deux modèles constitue un programme de recherches en soi, pour lequel nous proposons deux pistes d'investigation : la relation de la raison graphique à la raison computationnelle, et l'examen de chacun de deux modèles à l'aune de la catégorie de *représentation* comme catégorie tierce.

**Références**

Albèra, Ph. (1995). *Entretien avec Claude Helffer*. Lausanne: Contrechamps.

Bonardi, A. (2000). *Contribution à l'établissement d'un genre : l'opéra virtuel interactif*. Thèse de doctorat, Université Paris-IV.

Bonardi, A., Rousseaux, F. (2002). *Composing an Interactive Virtual Opera: The Virtualis Project*. Revue Leonardo, Journal of the International Society for the Arts, Sciences and Technology, volume 35, numéro 3, pp. 315-318.

Chion, M. (1983). *Guide des objets sonores*. Paris: Buchet/Chastel.

Dufourt, H. (2002). *About Musical creativity*. Lecture of the European Society for the Cognitive Sciences of Music. Liège, 5-8 avril.

Fingerhut, M. (1999). *The IRCAM Multimedia Library: a Digital Music Library*. IEEE Forum on Research and Technology Advances in Digital Libraries (IEEE ADL'99). Baltimore, MD (USA), 19-21 mai.

Goody, J. (1987). *The Interface between the Written and the Oral*. New York: Cambridge University Press.

Grabocz, M. (1999). *Méthodes d'analyse concernant la forme sonate.* Méthodes nouvelles – Musiques nouvelles – Musicologie et création, pp. 109-135. Strasbourg: Presses Universitaires de Strasbourg.

Jouvelot, P. (1998). *Music Composer's WorkBench : vers un environnement intelligent de l'enseignement de la composition tonale*. Journées d'Informatique Musicale, La Londe-les-Maures. Publications du Laboratoire de Mécanique et d'Acoustique du CNRS n°148, pp. A1 1-9.

Lerdahl, F., Jackendoff, R. (1983). *A generative theory of tonal music*. Cambridge: The MIT Press.

Pachet, F. (1994). *The MusES System: an Environment for Experimenting with Knowledge Representation Techniques in Tonal Harmony*. First Brazilian Symposium on Computer Music, SBCM'94, Caxambu, Minas Gerais, Brésil, pp. 195-201.

Rastier, F. (1994). *Sémantique pour l'analyse*. Paris: Masson.

Rousseaux, F., Bonardi, A. (2002). *Vagabonds, pédants ou philistins : choisir en beauté*. Art lyrique et art numérique. A propos d'une scénographie virtuelle interactive de Norma de Bellini, série " Conférences et Séminaires ", Observatoire Musical Français n°13, pp. 43-50. Paris: Université de Paris-Sorbonne.

Rink, J. (2002). *The Practice of Performance: Studies in Musical Interpretation*. Cambridge: Cambridge University Press.

Schaeffer, P. (1966). *Traité des objets musicaux*. Paris: Seuil.

Szendy, P. (2001). *Ecoute, une histoire de nos oreilles*. Paris: Editions de Minuit.

Xenakis, I. (1981). *Musiques formelles*. Paris: Stock.

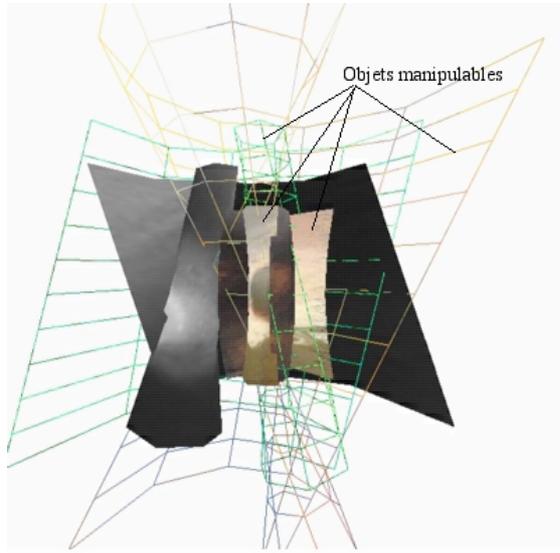

**Figure 1.** *Exemple d'objets 3D associés à des contenus musicaux dans Virtualis (source : Alain Bonardi)*

| Approche | Calcul | Réseaux de connaissances | Rapport charnel à la perception et à la compréhension | Raison computationnelle |
|---|---|---|---|---|
| Signe | Excès de **l'inscription** | Médiation de **la classification** | Fusion entre **écoute** et **manipulation** | Investissement d'esprit des processus de **répétition** machinique |
| Modalité | Calcul sur le signal numérique | Représentation des connaissances | Pratique corporelle d'une écoute/création et de sa transmission | Simulation comportementale |
| Ordres pertinents | Deux ordres sur les descripteurs : <br>- un ordre de dépendance du calcul <br>- un ordre de rentabilité économique | Acquérir les connaissances et les ordonner sont les deux faces d'une même activité de modélisation, qui a bien du mal à choisir entre des ordres théoriques (l'abstraction ou la généralisation) et des ordres pratiques (utilité pour un usager), d'où des problèmes d'acquisition et de maintenance des ontologies | Deux ordres pour classer les systèmes : <br>- potentiel de transformation offert par les contenus et les logiciels " players " <br>- degré d'interopérabilité des supports technologiques | Deux ordres pour classer les situations : <br>- la confiance de l'utilisateur <br>- la productivité des systèmes homme-machine |
| Lieux de développement | Le consortium de normalisation MPEG | Les communautés universitaires se réclamant de " l'acquisition des connaissances " | Les réseaux de partage/piratage de fichiers musicaux | Les laboratoires de conception des baladeurs musicaux numériques à disque dur |
| Exemple musical | La " durée effective " d'un signal sonore est un descripteur MPEG en voie de normalisation | Dans son *Traité des objets musicaux*, Pierre Schaeffer a proposé une ontologie pour la description des sons sur le mode de l'écoute réduite (au sens de la phénoménologie husserlienne) | Le " music-ripping " | Les logiciels de confection de " playlists " |